\documentclass[aps,prl,twocolumn,groupedaddress,showpacs]{revtex4}
\usepackage{graphicx}
\usepackage{dcolumn}
\usepackage{bm}
\usepackage{epsfig}
\usepackage{float}
\begin{document}





\title{3D Molecular dynamics simulations using spheropolytopes}

\author{S. A. Galindo-Torres }
\email{s.galinotorres@uq.edu.au}
\altaffiliation[Also at ]{Physics Department, Grupo de Simulacion de Sistemas Fisicos. Universidad Nacional de Colombia.}
\author{F. Alonso-Marroqu\'{\i}n}
\email{fernando@esscc.uq.edu.au}
\affiliation{ School of Physical Sciences,
               The University of Queensland, Qld. 4068, Brisbane, Australia}
\author{YC. Wang}
\affiliation{ ESSCC, The University of Queensland, Qld. 4068, Brisbane, Australia}

\date{\today}

\begin{abstract}
We present a simple and efficient method to simulate three-dimensional, complex-shaped,
interacting bodies. The particle shape is represented by Minkowski operators.  A
time-continuous interaction between these bodies is derived using simple concepts of
computational geometry. The model (particles + interactions) is efficient, accurate and easy
to implement, and it complies with the conservation laws of physics. 3D simulations of hopper
flow shows that the non-convexity of the particles strongly effects the jamming on granular flow.
\end{abstract}
\maketitle

Molecular Dynamics, MD is the art of modeling complex systems as a collection of particles
interacting each other. MD in three-dimensions using arbitrary particle shape is
a fundamental problem in several disciplines:
Drug molecules often act as a key in a lock formed by a protein
cavity, so that they can be designed using MD simulations \cite{carlson00};
Liquid crystals consisting of complex-shaped molecules exhibit
transition to a nematic phase, which can be investigated from particle-based
simulations of complex shaped objects\cite{pelzl99};  The large scale modeling of
geological materials in foundations, landslides and fault zones requires a
constitutive equation which can be constructed using MD-like models
~\cite{alonso04c}; Computing the motion of rigid and articulated bodies can
lead to new advances in robotics, automation \cite{mirtich98} and virtual
reality applications \cite{ruspini01}.

The most typical approach for these applications is to solve the dynamics of
interacting  rigid bodies, where their real shapes are approximated by
polyhedra \cite{mirtich98,baraff93,hasegawa04}.  The most difficult aspect for
the simulations is to model contact interactions. Contact force methods have been
proposed for two-dimensional (2D) models using polygons\cite{alonso04c,poeschel04}.
However, the extension of this method to three-dimensional (3D) simulations has proven
to be extremely difficult. In the simple case of  convex polygons, the force
is calculated as a function of their overlap area \cite{alonso04c}. However, the assumption
that elastic force is a function of their overlap leads to a non-conservative elastic
interaction \cite{poeschel04}.  An alternative approach is to assume that the  potential
elastic energy is a function of the overlap. Then  forces and torques are derived from this
potential~\cite{poeschel04}.  Both approaches have still not been extended to 3D, because the
calculation of the overlap between two polyhedra is computationally very expensive. This is
the main reason  why most of the commercial codes for particulate systems are  still based on
simulations with spheres, or clumps of spheres representing complex shaped objects \cite{lu07}.

An alternative solution for the 2D simulations of complex shaped particles has been proposed
recently \cite{alonso08epl}. The method introduces the concept of spheropolygons, which is the
object resulting from dilating a polygon by an sphere. The method not only guarantees energy balance
but also proves to be much more efficient that previous models to represent complex particle shape
\cite{alonso08b,galindo08a}. In 3D models, the dilation of a polyhedra by a sphere has a precise mathematical
meaning using the Minkowski operator. In our knowledge, Liebling and Pournin were the first to
introduce the Minkowski operators in particle-based simulations \cite{pournin05a,pournin05b}.
In order to calculate the  interactions, they assumed a single contact point between
the particles \cite{pournin05a}. These approach, however, leads to forces discontinuous in time and numerical
artifacts such as abrupt creation of mechanical energy. An alternative approach is proposed
by Pournin by calculating the overlap area between the particles \cite{pournin05c}. This approach in practice
is not feasible due to the high complexity of the boundary of the spheropolyhedra.
\begin{figure}[H]
\begin{center}
\includegraphics[trim=0cm 7cm 0cm 7cm,clip,scale=0.4]{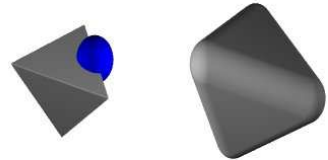}
\caption{The spherotetrahedron (right) is obtained by sweeping an sphere into a tetrahedron (left).}
\label{fig:dilation}
\end{center}
\end{figure}
In this Letter we present a solution to this problem by using the concept of spheropolytopes. They are generated from the Minkowski addition of a polytope by an sphere,  which is nothing more
than the object resulting from sweeping a sphere around a polytope. A polytope is a generic
mathematical concepts that can refer to polygons, polyhedra or polygonal curves in 3D. The
polytope is regarded as a collection of {\it features} in the three-dimensional Euclidian space:
vertices, edges and faces. The interaction between two spheropolytopes is  calculated as
a function of the distance between their features. The molecular dynamics is implemented in
a simple, efficient and  elegant algorithm,  which complies with conservation laws
of physics. We believe that this model will lead to a wide range of applications of molecular
dynamics, as complex particle shape and realistic interactions can be captured in
a unified framework using well established concepts of molecular dynamics and computational
geometry.

\begin{figure}[H]
\begin{center}
\includegraphics[trim=0cm 8cm 0cm 8cm,clip,scale=0.4]{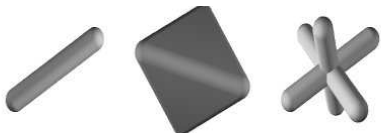}
\caption{Spheropolytopes generated  as sphere $\oplus$ line segment ({\it rice}), sphere $\oplus$ tetrahedron ({ \it tetra})
and sphere $\oplus$ polyline ({\it yermis}).}
\label{fig:spheropolytopes}
\end{center}
\end{figure}

For the representation of arbitrary shaped particles we introduce the mathematical
concept of Minkowski sum. Given two sets of points P and Q in an Euclidean space, their
Minkowski sum is given  by $P \oplus Q =\{\vec x+ \vec y~|~ \vec x \in P,~ \vec y \in Q \}$.
This operation is geometrically equivalent to the sweeping of one set around
the profile of the other without changing the relative orientation (Fig \ref{fig:dilation}).  Examples
of Minkowski sums are the spheropolygons (sphere $\oplus$ polygons) \cite{alonso08epl}, spherocyllinder
(sphere $\oplus$ line segment) \cite{pournin05a}, the spherosimplex  (sphere $\oplus$ simplex)
\cite{pournin05b} and the spheropolyhedron (sphere  $\oplus$ polyhedron) \cite{pournin05c},
All these objects can be enclosed in the generic shape of spheropolytopes, which consists of a set of
vertices ${V_i}$, edges ${E_i}$, polygonal faces ${F_i}$ and the radius $r$ of the
sweeping sphere wich will be called sphero-radius. The three examples of spheropolytopes
we will consider in this letter is the  {\it rice}, the  {\it tetra} and  the {\it yermis} in
the Fig. \ref{fig:spheropolytopes}.
For the calculation of the contact force between spheropolytopes, we require expressions for the distance
between their features. Given two features $G_i$ and $G_j$ of the spheropolytopes $i$ and
$j$, their distance is  defined as $d(G_i,G_j)=||\vec X_i - \vec X_j ||$, where
$\vec X_i$ and $\vec X_j$ is the closest points belonging to either feature.
We start with the formula of the distance between a vertex $V_i$ and an edge $E_j$.
Let's consider the line $\ell_i$ containing the edge. First we calculate the closest point of
this line to the vertex. if the point lies on the edge this is the closest point of
the edge to the vertex. Otherwise we take the minimal of the distance from the vertex to
both endings of the edge.

Next we consider the distance  between two edges $E_i$ and $E_j$. Let's the lines
$\ell_i$ and $\ell_j$ containing the edges, we find the two points on these lines
whose distance is minimal. If both points belong to the edges, they define the minimal
distance between the edges. Otherwise the distance between the two edges is calculated
as the minimal of the distance between each vertex of one edge and the other edge.

For the calculation of the distance between a vertex $V_i$ and and face $F_j$ we consider
the plane $\Pi_j$ containing the face.  First we project the point on the plane. If the
projection of the point in the plane lies inside the polygon face then the distance is
calculated between these two points. If the projection lies outside the polygon then the
distance  is calculated to the closest point in the polygon boundary.

These formulas of distance are used to calculate the force between two spheropolytopes.
the force $\vec{F}_{ij}$ on the i-spheropolytope by the j-spheropolytope is taken
as a superposition of the interaction between each pair  of edges $\vec{F}(E_i,E_j)$
and each pair of vertex-face $\vec{F}(V_i,F_j)$ for the spheropolytope pair,

\begin{equation}
\label{eq:totalf}
\vec{F}_{ij}=\sum_{E_i,E_j}\vec{F}(E_i,E_j)+\sum_{V_i,F_j}\vec{F}(V_i,F_j)+\sum_{V_j,F_i}\vec{F}(V_j,F_i).
\end{equation}

\noindent
The force $F(G_i,G_j)$ associated to the two features (edge-edge or vertex-face) is assumed to depend on the
overlapping length $\delta$ between them

\begin{equation}
\label{eq:deltaee}
\delta(G_i,G_j)=R_i+R_j-d(G_i,G_j),
\end{equation}

\noindent
with $d(G_i,G_j)$ the distance between the features of the spheropolytopes $R_i$ the spheroradius of the
i-th spheropolytope.  The point of contact between the two features is calculated by taken the spheres
of radius $R_i$ and $R_j$ centered in the closest points $\vec X_i$ and $\vec X_j$, and finding the intersection between the line connecting these two points and the line connecting the two intersection points of the spheres.
This contact point results as

\begin{equation}
\label{eq:rfee}
\vec{R}(G_i,G_j)=\vec{X}_i+\frac{R_i^2-R_j^2+d^2(G_i,G_j)}{2d(G_i,G_j)}\vec{n},
\end{equation}

\noindent
where $\vec n = \frac{\vec X_j - \vec X_i}{||\vec X_j - \vec X_i||}$.
From the point of application of the contact forces we get the torque on the
i-spheropolytope by the  j-spheropolytope:

\begin{equation}
\label{eq:totalf}
\vec{\tau}_{ij}=\sum_{E_i,E_j}\vec{\tau}(E_i,E_j)+\sum_{V_i,F_j}\vec{\tau}(V_i,F_j)+\sum_{V_j,F_i}\vec{\tau}(V_j,F_i).
\end{equation}

\noindent
where

\begin{equation}
\label{eq:torque}
\vec{\tau_i}(G_i,G_j)=(\vec{R}(G_i,G_j)-\vec{r}_i)\times\vec{F}(G_i,G_j)
\end{equation}
with $\vec{r}_i$ the center of mass of the i-spheropolyhedron.

Since the formulas of distance are continuous functions on the degrees of freedom of
the spheropolytopes, the total force is continuous too. This avoids the problems of
discontinuity in time of the forces in previous models \cite{mirtich98}. Different forces can be included
in this model: for example, a force derived from a potential function of the distance leads
to a conservative systems; forces depending on the relative velocities at the
contact points leads to dissipative granular materials; Forces depending on the history of
relative velocity at the contacts represent frictional granular systems; more sophisticated
forces can be used to simulate biomolecules. The electrostatic interaction between the
molecules can be modeled by allowing the force depend on the closest points between the
features.

%
%
%
%
%
%

\begin{figure}[H]
\begin{center}
\includegraphics[scale=0.3]{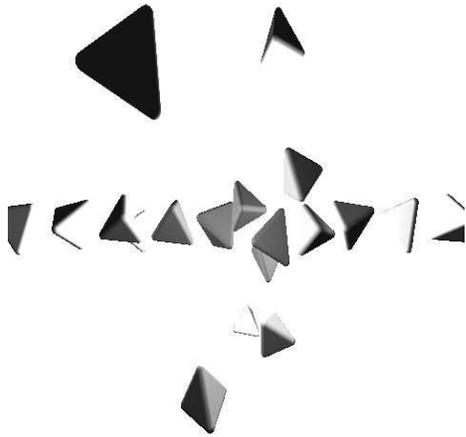}
\caption{Spherotetrahedra collider for checking the conservation laws.}
\label{fig:tetracollider}
\end{center}
\end{figure}

The first numerical experiment presented in this paper is the spheropolytopes collider (SPC).
The simulation consists of colliding opposing beams of spheropolytopes. We consider the three
shapes shown in Fig. \ref{fig:spheropolytopes} . The SPC is build with the intention to produce enough
collisions to test the basic conservative laws of physics: the conservation of energy, linear
and angular momentum. The experimental setup is showed in Fig. \ref{fig:tetracollider}, two rows
of spherotetrahedra are set in a collision course with random orientation and angular velocity. The
contact force between features is calculated as

\begin{equation}
\label{eq:fvf}
\vec{F}(G_i,G_j)= - k_n \delta(G_i,G_j) \vec{n},
\end{equation}

\noindent
where $k_n$ is the elastic constant. Once all the forces are calculated we integrate Newton's second law using the Verlet algorithm for the translation coordinates. The Euler equations form angular momentum is integrated using and the Fincham Leap Frog algorithm, based on the quaternion formalism, for the orientation coordinates \cite{wang2006ips}. The Table \ref{tab:conservation} shows the percentage error of the three conservation laws (energy, linear and angular momentum) for two different time steps. The consistency of our numerical method is verify as the error decreases as the time step is smaller. The discretization error is larger in the yermis-particles, because they produce much more collisions than the other particles.
\begin{figure}[H]
  \begin{center}
    \epsfig{file=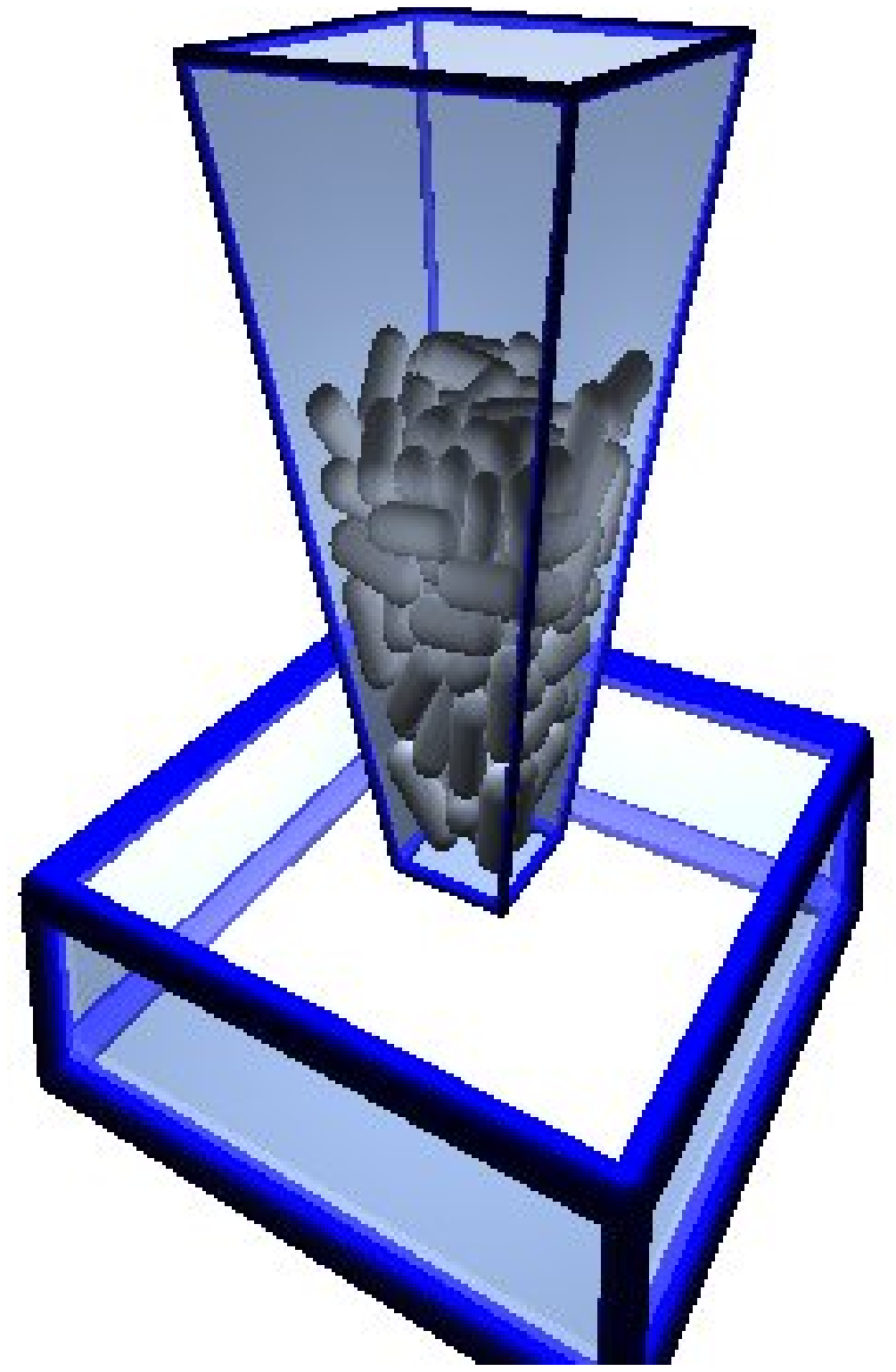,width=0.45\linewidth}
    \epsfig{file=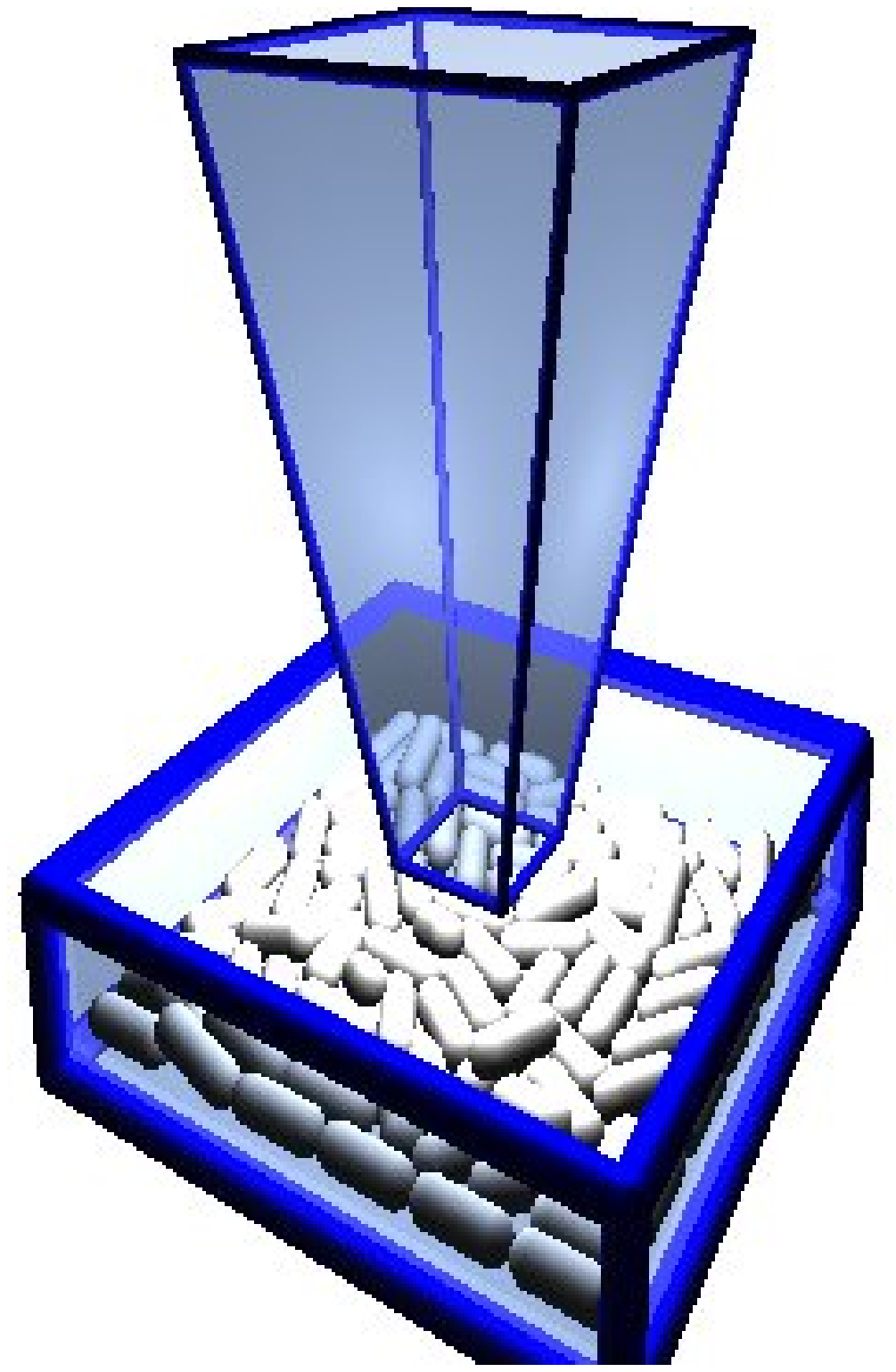,width=0.45\linewidth} \\~\\
    \epsfig{file=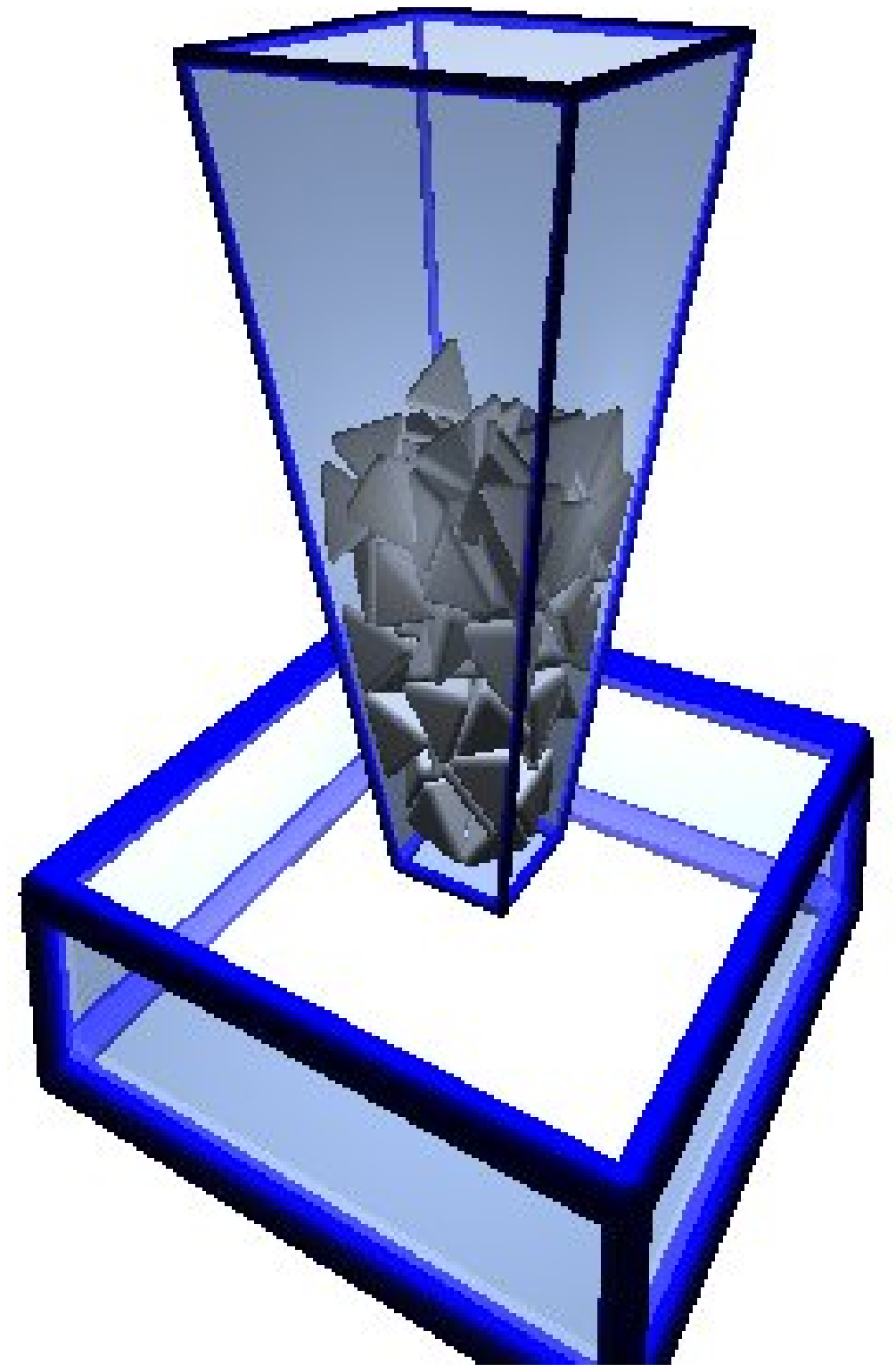,width=0.45\linewidth}
    \epsfig{file=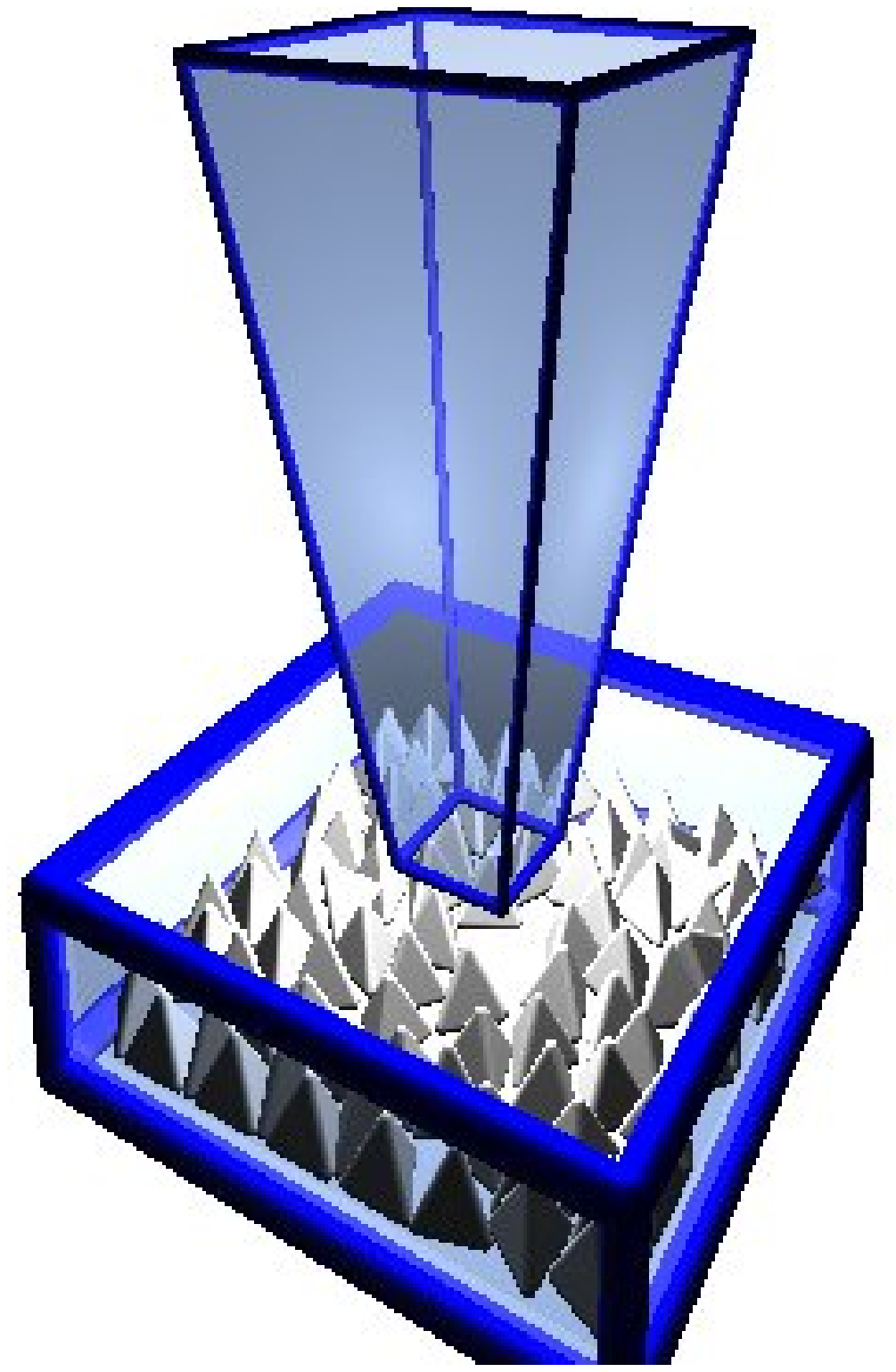,width=0.45\linewidth} \\~\\
    \epsfig{file=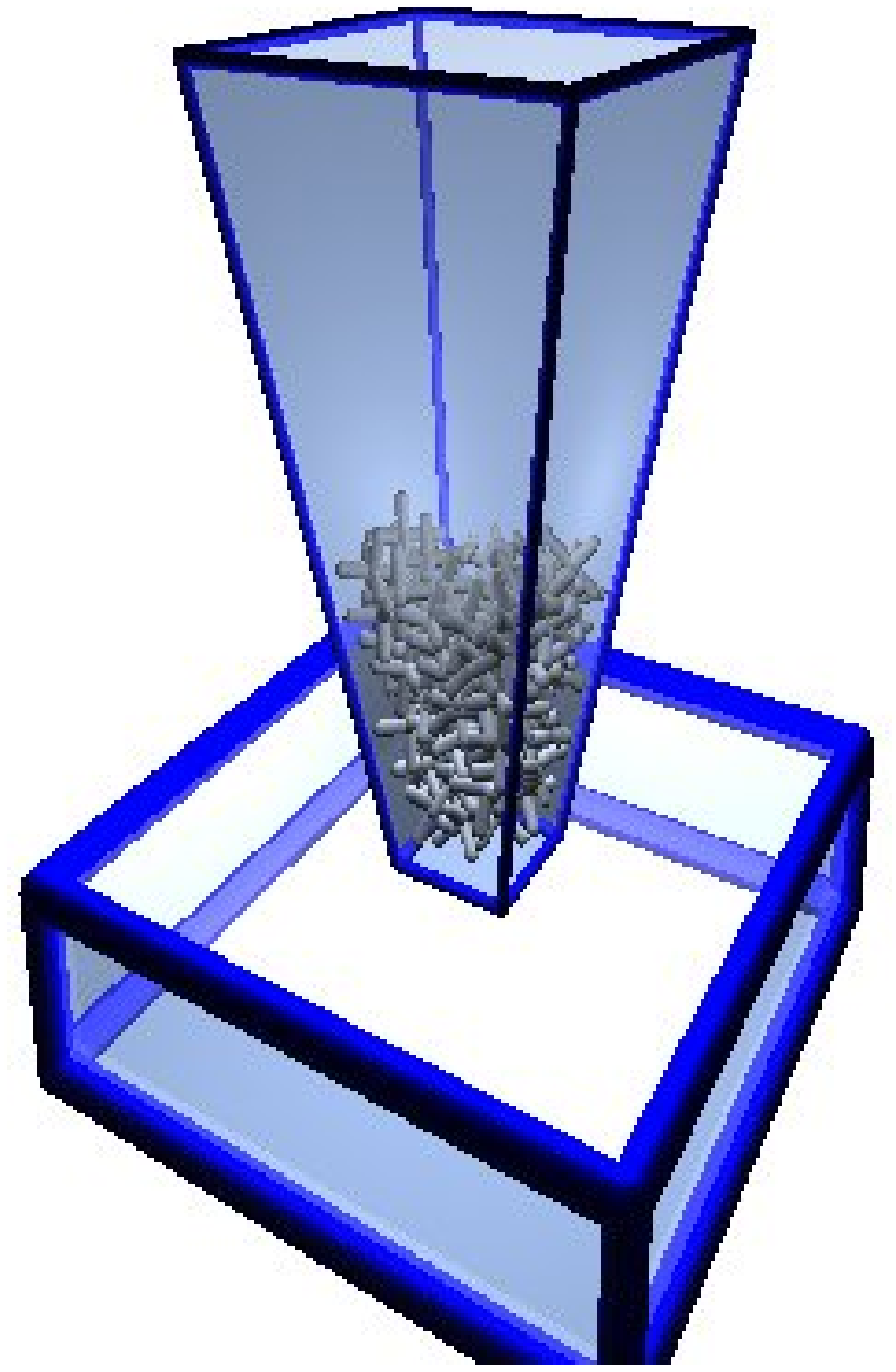,width=0.45\linewidth}
    \epsfig{file=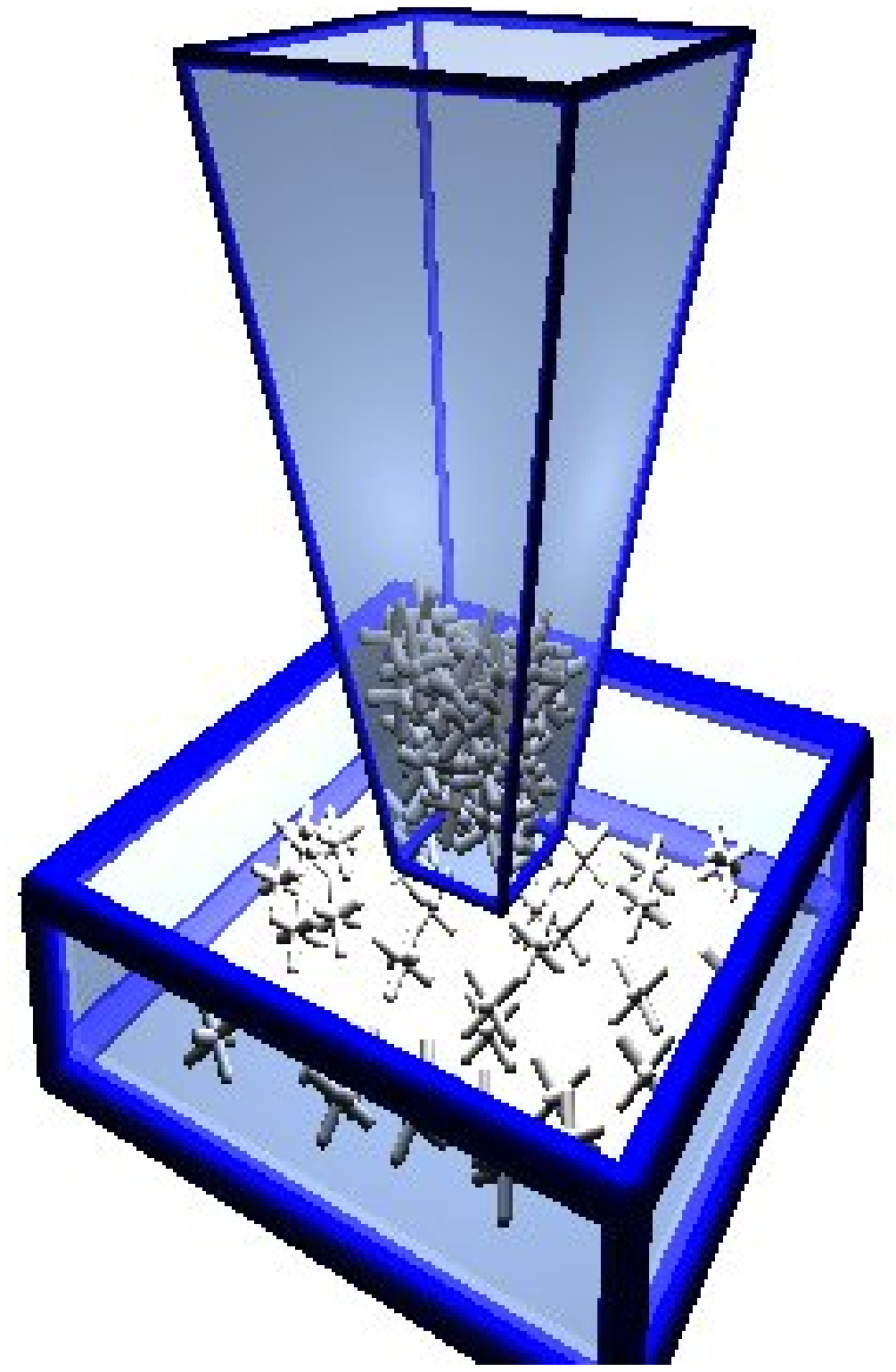,width=0.45\linewidth}

   \caption{Hopper flow simulations at initial (left) and final (right) stages for rice (above), tetra (center) and yermis (below). The parameters are the same than in the SPC experiment. The coefficient of viscosity $\gamma$ is equal to 0.2 $s^{-1}$}
   \label{fig:hopper}
  \end{center}
\end{figure}
\begin{table}
\begin{ruledtabular}
\begin{tabular}{lccc}
Particle & $\frac{\Delta E}{E}$ & $\frac{\Delta p}{p}$ & $\frac{\Delta L}{L}$ \\
rice $\Delta t=10^{-4}$ & $3.6 \times 10^{-3}$ & $2.1 \times 10^{-9}$ & $2.0 \times 10^{-4}$ \\
rice $\Delta t=10^{-5}$ & $4.0 \times 10^{-4}$ & $1.1 \times 10^{-10}$ & $5.8 \times 10^{-5}$ \\
tetra $\Delta t=10^{-4}$ & $2.2 \times 10^{-3}$ & $3.2 \times 10^{-9}$ & $4.6 \times 10^{-5}$ \\
tetra $\Delta t=10^{-5}$ & $2.0 \times 10^{-4}$ & $5.3 \times 10^{-10}$ & $1.5 \times 10^{-6}$ \\
yermis $\Delta t=10^{-4}$ & $3.1 \times 10^{-2}$ & $1.9 \times 10^{-9}$ & $3.5 \times 10^{-3}$ \\
yermis $\Delta t=10^{-5}$ & $2.6 \times 10^{-3}$ & $1.8 \times 10^{-10}$ & $2.3 \times 10^{-4}$ \\
\end{tabular}
\end{ruledtabular}
\caption{\label{tab:conservation} Percentage error in the numerical simulations calculated from the mechanical energy (E), angular momentum (L) and linear momentum(p) refereed to their initial values. The simulation time is 20 s, the mass of the particles is 1. kg and the stiffness is 10000 N/m}
\end{table}

The immediate extension of the model is to include visco-elastic forces:

\begin{equation}
\label{eq:fvf}
\vec{F}(G_i,G_j)= - k_n \delta(G_i,G_j) \vec{n} + \gamma \vec v_c
\end{equation}

where $\gamma$ is the viscocity force and $\vec v_c$ is the relative
velocity of the particles at the contact. This contact force offers
an interesting application of this model: the study of the effect of particle
shape on the jamming phenomenon of  granular flow. The flow may happen
when particles are  discharged through a small opening, but  particles may became
jammed when the  opening is smaller than a critical value.  Modeling of gravity
flow has  been done using circular or spherical  particles \cite{to2001jgf}, but the
effect of  shape on flow has not been thoughtfully investigated. In particular,
non-convex particles is expected to jam more easily than convex, or circular
particles.

\begin{figure}[H]
\begin{center}
\includegraphics[bb = 0 0 1140 1614,scale=0.25,trim=0cm 0cm 15cm 35cm,clip]{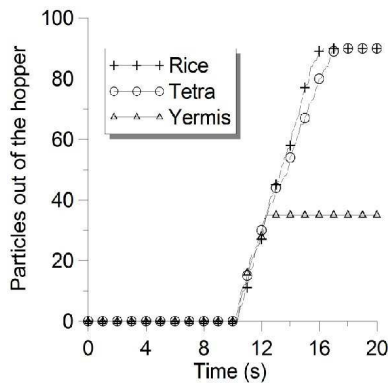}
\caption{Number of particles exiting the hopper in the setup shown in Fig \ref{fig:hopper}. The simulation consider 90 particles. Note that in the {\it yermis} case the flow stops at 12 s due to clogging.}
\label{fig:hopperflow}
\end{center}
\end{figure}

Granular flow with convex and non-convex particles is presented using
the same three particles geometries shown in Fig. \ref{fig:spheropolytopes}.
The simplicity of our model allow us to represent the hopper and the container as just another example of spheropolytopes, see Fig. \ref{fig:hopper}. Contrary to previous findings \cite{poeschel04} our convex shaped particles do not become clogged in the hopper. This is because we have not introduced an static frictional force yet. However, as an striking result, the non-convex particles get stuck without static friction, as can be seen in Fig. \ref{fig:hopperflow}.

Modeling interacting particles using spheropolytopes has
several advantages  with respect to other existing particle-based models:
i) The possibility to model non-convex particles (in our case {\it yermis}, hoppers and containers); ii) a realistic representation of the surface
curvature of particles; iii) guaranteed compliance with physical laws; iv) numerical consistency, guaranteed by the continuity in the proposed contact law.
v) efficiency, given by a simple model for the contact law relying only on distance calculations. This is radically different from previous approaches where the contact forces are calculated in base of overlaps \cite{mirtich98,pournin05c}

The most interesting aspect of this model is to provide a general framework
for generic particle shape and contact interactions.  Spheropolytopes is a
very general shape which can be uses to represent biomolecules, polymers,
rocks, meteors, etc. For the modeling of geological materials, particles
with random shapes and tunable roundness can be generated by applying
Minkowski operators on Voronoi diagrams \cite{galindo08a}. Comminution
processes can be model by solving the continuum stress equation of a given
spheropolyhedra and originate from it fracture planes and hence secondary
spheropolyhedra.

This work is supported by the Australian Research Council (project number DP0772409 )
and the UQ Early Carrier Research Grant.

\end{document}